\pdfoutput=1
\documentclass[runningheads]{llncs}
\usepackage[T1]{fontenc}
\usepackage{graphicx}
\usepackage{tabularx}
\usepackage{multirow}
\usepackage{footnote}
\usepackage{hyperref}
\usepackage{amssymb}
\usepackage{amsmath}
\hypersetup{
  colorlinks = true,
  linkcolor  = blue,
  urlcolor   = blue,
  citecolor  = black,
}
\usepackage{url}
\usepackage{cleveref}
\newcommand*{\fullref}[1]{\hyperref[{#1}]{\cref*{#1} \nameref*{#1}}}
\newcommand*{\Fullref}[1]{\hyperref[{#1}]{\Cref*{#1} \nameref*{#1}}}
\newcommand*{\secref}[1]{\hyperref[{#1}]{\autoref*{#1}}}
\newcommand*{\Secref}[1]{\hyperref[{#1}]{\Cref*{#1}}}

\let\oldFootnote\footnote
\newcommand\nextToken\relax

\renewcommand\footnote[1]{%
    \oldFootnote{#1}\futurelet\nextToken\isFootnote}

\newcommand\isFootnote{%
    \ifx\footnote\nextToken\textsuperscript{,}\fi}

\begin{document}
\title{SciCom Wiki}
\subtitle{Fact-Checking and FAIR Knowledge Distribution \\for Scientific Videos and Podcasts}
\titlerunning{SciCom Wiki}
%
\author{Tim Wittenborg\inst{1, 2}\orcidID{0009-0000-9933-8922} 
\and Constantin Sebastian Tremel\orcidID{0009-0008-4111-4115} 
\and Niklas Stehr 
\and Oliver Karras\inst{3}\orcidID{0000-0001-5336-6899} 
\and Markus Stocker\inst{3}\orcidID{0000-0001-5492-3212} 
\and Sören Auer\inst{1,2,3}\orcidID{0000-0002-0698-2864}
}
\authorrunning{Wittenborg et al.}
%
\institute{L3S Research Center, Leibniz University Hannover, Hannover, Germany\\
\email{\{tim.wittenborg, soeren.auer\}@l3s.de} \and Cluster of Excellence SE²A – Sustainable and Energy-Efficient Aviation, Technische Universität
Braunschweig, Germany \and
TIB - Leibniz Information Centre for Science and Technology, Hannover, Germany\\
\email{\{oliver.karras, markus.stocker, soeren.auer\}@tib.eu}}

\maketitle              
\begin{abstract}
Democratic societies need accessible, reliable information.
Videos and Podcasts have established themselves as the medium of choice for civic dissemination, but also as carriers of misinformation.
The emerging Science Communication Knowledge Infrastructure (SciCom KI) curating these increasingly non-textual media is still fragmented and not adequately equipped to scale against the content flood.
Our work sets out to support the SciCom KI with a central, collaborative platform, the SciCom Wiki, to facilitate FAIR (findable, accessible, interoperable, reusable) media representation and thereby the fact-checking of their content, particularly for videos and podcasts.
Building an open-source service system centered around Wikibase, we survey requirements from 53 stakeholders, individually refine these in 11 interviews, and evaluate our prototype based on these requirements with another 14 participants.
To address the most requested feature, fact-checking, we developed a neurosymbolic computational fact-checking approach, converting heterogenous media into knowledge graphs.
This increases machine-readability and allows comparing statements against equally represented ground-truth.
Our computational fact-checking tool was iteratively evaluated through 10 expert interviews, while a public user survey with 43 participants verified the necessity and usability of our tool.
Overall, our findings identified several needs to systematically support the SciCom KI.
The SciCom Wiki, as a FAIR digital library complementing our neurosymbolic computational fact-checking framework, was found suitable to address the raised requirements.
Further, we identified that the SciCom KI is severely underdeveloped regarding FAIR knowledge and related systems facilitating its collaborative creation and curation.
Our system can provide a central knowledge node similar to Wikidata, yet a collaborative effort is required to scale the required features against the imminent (mis-)information flood.

\end{abstract}
\section{Introduction}
Over 100 zettabytes of online web content are created, captured, copied, and consumed globally every year~\cite{statista_daten_2023}.
Every minute, humanity watches 43 years of streaming content and sends over 360 thousand tweets~\cite{domo}.
The accuracy of this content is often unclear to consumers, particularly where platforms lack mechanisms such as outlinks, user comments, public discussions, or fact-checking tools ~\cite{brennen_types_2020}.
Yet, accurate information is particularly important for political topics like climate action, where a rapidly closing window of opportunity~\cite{lee_ipcc_2023} meets insufficient policies~\cite{emissiongapReport2023} and debates not aligning with the overwhelming scientific consensus~\cite{lee_ipcc_2023,cook2013}.
While Knowledge Infrastructures (KIs) like Wikipedia, Wikidata, and Wikisource successfully curate textual information at scale~\cite{giles_internet_2005}, even their ressources and scalability are limited.
Their broad scope and high notability criteria exclude a wide range of scientifically relevant media, resulting in structural gaps.
Wikimedia has identified these gaps and started the Wikibase Ecosystem\footnote{\url{https://wikiba.se/}}~\cite{varvantakis_wikibase_2025}, sprouting hundreds of dedicated knowledge bases, like MiMoText~\cite{schoch_smart_2022} for literary history and literary historiography.
Despite this, modern non-textual media (NTM) such as video and audio remain largely unaddressed.

We present a our work towards a FAIR knowledge platform for the Science Communication  KI, as illustrated in \secref{fig:overview}.
Our contributions include:
\begin{figure}[b]
   \vspace{-0.5cm}
    \centering
    \includegraphics[width=1.0\linewidth]{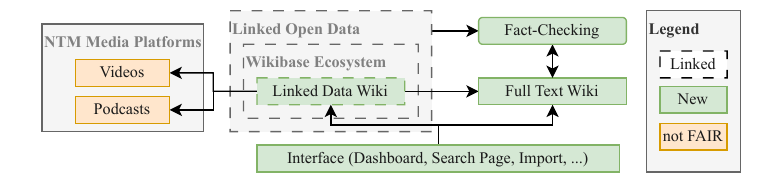}
    \caption{Extension of the Science Communication KI (SciCom KI) with our systems.}
    \label{fig:overview}
\end{figure}

\textbf{1.)} A digital library and interface to provide tool- and infrastructure-support for collaborative media representation, processing, and annotation, as detailed in the master thesis by Stehr~\cite{stehr_digitale_2025}.
Our system builds on a microservice architecture, open source software, and the FAIR data principles~\cite{wilkinson_fair_2016}.
Requirements were gathered via 53 anonymous survey participants and 11 stakeholder interviews, while the final implementation was evaluated by 14 stakeholders, confirming a good user experience.
While the system proved capable of meeting several requirements already, extensive future work is required to modularly adress all.

\textbf{2.)} A streamlined workflow for fact-checking content -- the most requested SciCom KI aspect -- which semantifies both ground-truth sources and popular media to comparatively score scientific accuracy, as detailed in the master thesis by Tremel~\cite{tremel_scientific_2024}.
The open-source implementation was evaluated via an anonymous user survey with 43 participants and 10 expert interviews,
validating both its usability and relevance, while also highlighting the need for further SciCom KI development to scale fact-checking amid the information flood sustainably.

This work is structured as follows: \Secref{sec:related} presents the background and related work focusing on SciCom, KI, and fact-checking.
\Secref{sec:sciMediaKI} details the approach, implementation, and evaluation of the digital library, particularly highlighting the need for systematic fact-checking support.
We meet this need in \secref{sec:fact-checking}, equally illustrating the approach and evaluation.
The resulting findings are discussed in \secref{sec:discussion}, detailing feature work and concluding in \secref{sec:conclusion}. 

\section{Background and Related Work\label{sec:related}}
We summarize the fundamentals of SciCom KI to then present its state of the art in non-textual media processing and computational fact checking, particularly in the context of a climate communication use-case.

\paragraph{Knowledge Infrastructure}
Knowledge Infrastructures (KI) are defined by Edwards~\cite{10.5555/1805940} as ``\textit{robust networks of people, artifacts, and institutions that generate, share, and maintain specific knowledge about the human and natural worlds}''.
Notable examples include the Intergovernmental Panel on Climate Change (IPCC) and the Wikipedia Community, with its individual contributors, formal and informal organizations and (sub-)projects, digital libraries, tools, etc.
They are characterized by bringing together ``\textit{a diversity of actors, organizations and perspectives from, for instance, academia, industry, business and general public}''~\cite{karasti_knowledge_2016}, enduring beyond any individual project time~\cite{karasti_infrastructure_2010}.
As an emerging scientific field, it bridges the gap between different audiences and demographics, leading to a natural adjacency to another domain of open science: Science Communication.

\paragraph{Science Communication}
Burns et al.~\cite{burns_science_2003}
define Science Communication (SciCom) as ``\textit{the use of appropriate skills, media, activities, and dialogue to produce one or more of the following personal responses to science (the AEIOU vowel analogy): Awareness, Enjoyment, Interest, Opinion-forming, and Understanding}''.
It aims to improve public understanding, acceptance, trust, and support while providing a venue to gather broad feedback, local knowledge, and civic needs regarding valuable research aims and applications~\cite{kappel_why_2019}.
Videos and podcasts in particular are widely used as knowledge distribution media.
MacKenzie~\cite{mackenzie_science_2019} has curated 952 science podcasts, \href{https://wissenschaftspodcasts.de/podcasts/}{wissenschaftspodcasts.de} 386, the world lecture project\footnote{\url{https://world-lecture-project.org/}} 59,634 videos, the TIB-AV Portal~\cite{marin_arraiza_tibav_2015} 50,690.
All are actors in the SciCom KI.

\paragraph{Science Communication Knowledge Infrastructure}
The SciCom KI is hence the knowledge infrastructure dedicated to science communication, with its potentially fragmented individual actors including people, artifacts, and institutions. 
Notable contributors include the Science Media Center (SMC) Global Network\footnote{\url{https://www.sciencemediacentre.org/international-smcs/}}, their national chapters, and individual members.
Sharing and maintaining knowledge are inherent to SciCom KI, which utilizes textual and non-textual artifacts to derive, preserve, and convey knowledge~\cite{kulczycki_transformation_2013}.
Particularly videos are widely used and meticulously researched, with hundreds of different qualities and properties being analyzed and annotated~\cite{navarrete_closer_2023}.
This secondary data is usually stored in isolated repositories, disconnected from the original media, unless it is captured within FAIR infrastructure -- for instance, through licensing media in the TIB AV-Portal~\cite{marin_arraiza_tibav_2015} or by meeting the notability criteria\footnote{\url{https://www.wikidata.org/wiki/Wikidata:Notability}} of Wikidata~\cite{vrandecic_wikidata_2014}.
Yet, capturing this knowledge is especially relevant when assessing information that coexists within the same communication channels, but varies in quality, purpose, scope, evaluation standards, accessibility, and revenue generation~\cite{fahnrich_exploring_2023,hagenhoff_neue_2007}.
Whereas mechanisms such as peer reviews and narrow search index classification help protect internal scholarly communication from false claims, public discourse is far more susceptible to misinformation and requires consistent fact-checking.

\paragraph{Fact checking\label{sec:fc}}
Ciampaglia et al.~\cite{ciampaglia_computational_2015} state that ``\textit{fact checking by expert journalists cannot keep up with the enormous volume of information that is now generated online}''.
These experts often operate within organizations, such as Science Feedback or FactCheck.org, which coordinate guidelines and insights through overlapping meta networks such as the International Fact-Checking Network (IFCN) (141 members), European Digital Media Observatory (EDMO) (93), and the European Fact-Checking Standards Network (EFCSN) (55).
While all of these organizations are committed to specific standards for selecting, investigating, reviewing, and publicizing fact-checks, they vary significantly in the types of content they assess, their rating systems, and the degree of automation.

\paragraph{Computational fact checking\label{sec:cfc}}
Ciampaglia et al.~\cite{ciampaglia_computational_2015} propose computational fact-checking based on semantic proximity, aggregating the generality (degree) of two concept nodes in a knowledge graph by tracing the path in between.
Thorne et al.~\cite{thorne_fever_2018} introduce FEVER (Fact Extraction and VERification), a dataset for verification against textual sources.
Dessí et al.~\cite{dessi_scicero_2022} present an approach that extracts text from research articles to generate a knowledge graph of research entities automatically.
The reuse of top- and mid-level ontologies, such as the widely used Basic Formal Ontology (BFO) and Common Core Ontology (CCO), facilitates collective efforts and prevents the duplication of work~\cite{davarpanah_climate_2023}.

\paragraph{Ground truth knowledge graphs\label{sec:kg}}
Islam et al.~\cite{islam_knowurenvironment_2022} proposed a knowledge graph for climate change, extracting statements from 152 thousand domain-relevant scientific articles selected from an 8.1 million article dataset.
Of those 411,860 triples, 24,263 were identified as trusted unique triples and stored in a normalized, but not semantically disambiguated CSV\footnote{\url{https://github.com/saiful1105020/KnowUREnvironment/}}.
The sources used to construct such ground truth knowledge graphs must be carefully selected, i.e., reproducible, traceable, reputable, and peer-reviewed primary literature.
Despite collaborative efforts such as the IPCC, Climate Change Performance Index (CCPI), Corporate Climate Responsibility Monitor (CCRM), or Science Daily Climate Change (SciDCC) providing extensive knowledge synthesis, the outputs often culminate in PDF format, limiting all aspects of FAIRness~\cite{wilkinson_fair_2016}, especially interoperability.

Current infrastructure remains insufficient to manage the growing volume of (mis-)information.
Despite robust systems such as the Open Research Knowledge Graph (ORKG)~\cite{auer_improving_2020}, essential knowledge is still frequently locked in siloed formats, even in key KI such as the IPCC.
Establishing a reliable knowledge foundation requires a collective shift towards FAIR data, yet the SciCom KI in particular remains underdeveloped.
We seek to fill this gap in the SciCom KI by providing a digital library dedicated to key SciCom artifacts, videos and podcasts, to facilitate FAIR fact-checking at a collaboratively sustainable scale.

\section{SciCom Wiki: FAIR Non-Textual Media Representation\label{sec:sciMediaKI}}
To consolidate the fragmented SciCom KI efforts to curate NTM, we design, develop, and evaluate the SciCom Wiki, an open-source linked open data library for scientific videos and podcasts, detailed in the master thesis by Stehr \cite{stehr_digitale_2025}.

\subsection{Approach\label{sec:nik_approach}}

\paragraph{Mission Goal}
Develop a platform to index information on scientific videos and podcasts, to guarantee free access and participation for all users to create and curate content alongside the principles of open source and open access.

\begin{figure}[tb]
    \vspace{-0.5cm}
    
    \includegraphics[width=1.0\linewidth]{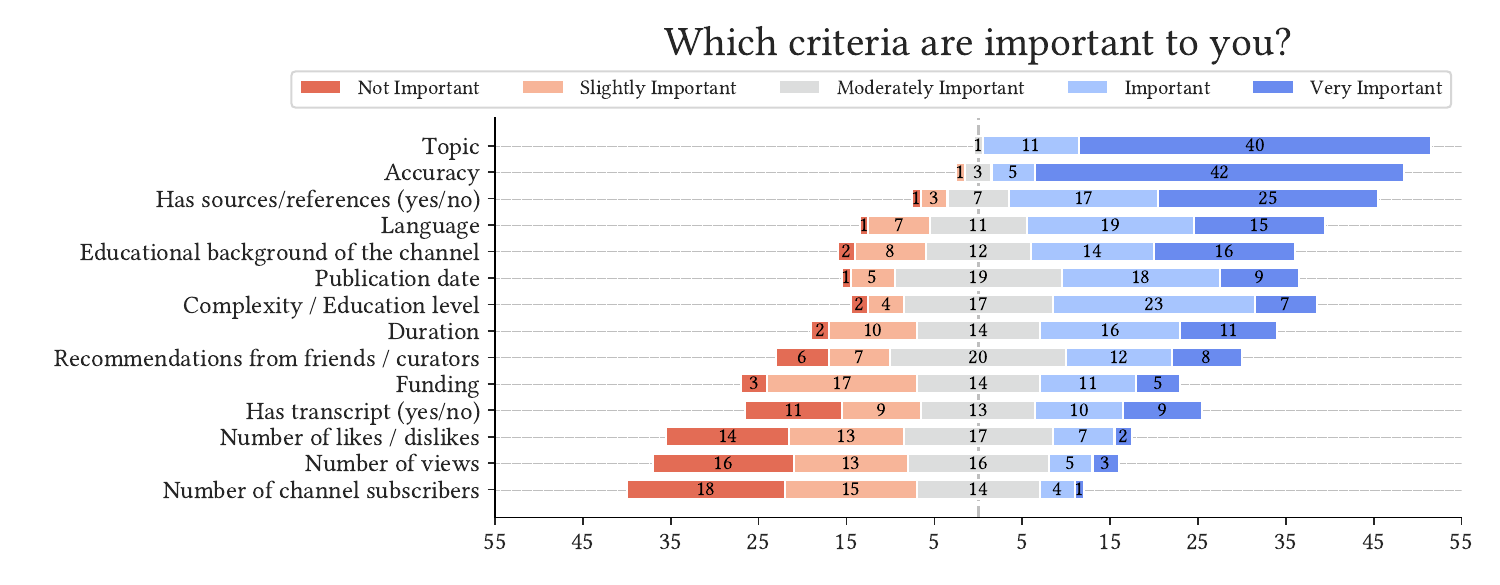}
    
    \vspace{-0.3cm}
    
    \includegraphics[width=1.0\linewidth]{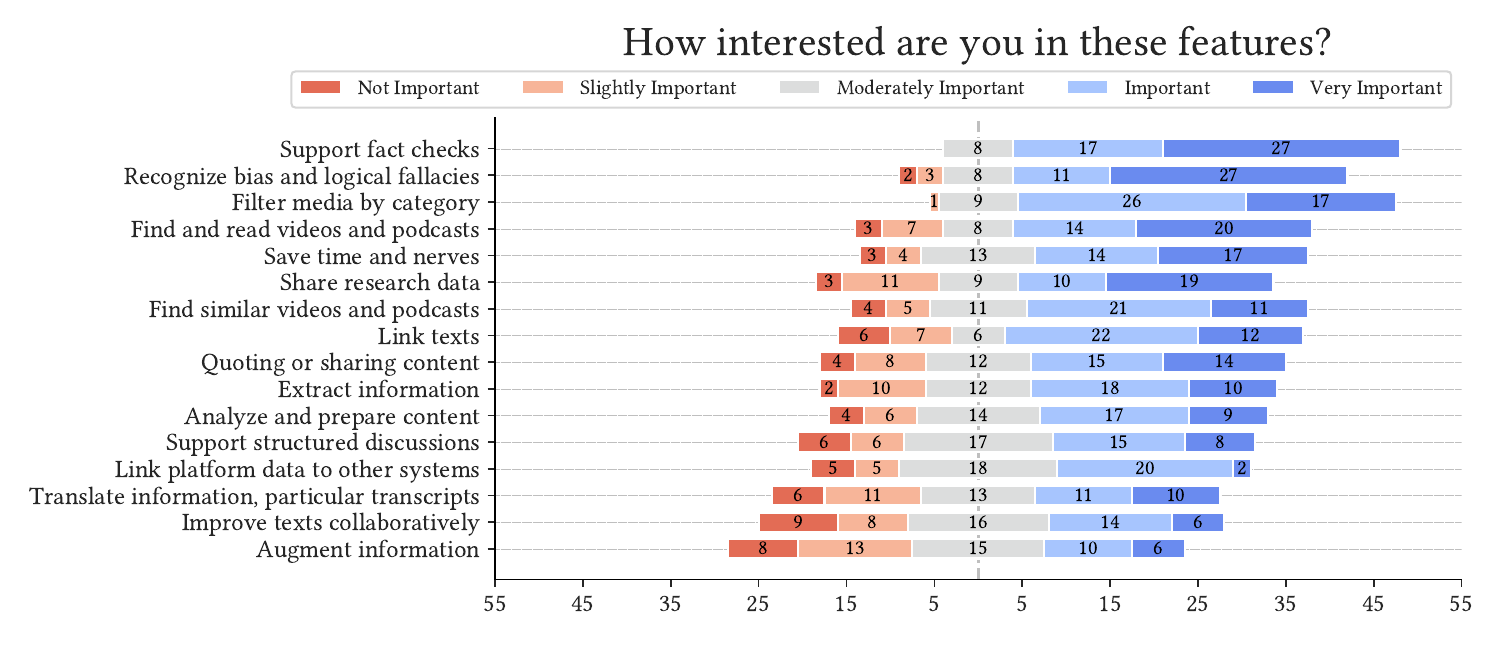}
    
    \vspace{-0.5cm}
    
    \caption{Results of 53 participants assessing the importance of Sci KI media criteria (above) and features (below), ranked by their importance averaged over all responses.}
    \label{fig:caf_ranking}
    
    \vspace{-0.5cm}
\end{figure}

\paragraph{Requirements Elicitation}
A domain analysis identified a set of intermediate requirements and six key stakeholder roles within the SciCom KI: viewer, researcher, teacher, content creator, curator, and developer.
These preliminary conceptions informed the survey design, which was distributed digitally, via flyer, and presented at the MediaWiki Users and Developers Conference~\cite{wittenborg_scicom_2024}.
The survey asked participants to indicate which roles they associate with when engaging with SciCom media, as well as what features they require to navigate the content flood.
Of the 53 participants, 40 were interested or very interested as viewers, 18 as developers, and 14 as researchers. 
Only seven were (very) interested as curators, another seven as teachers, and only four as content creators.
The survey and results are available online\footnote{\url{https://github.com/xEatos/survey-auswertung}}, most important of which are the criteria and feature rankings depicted in \secref{fig:caf_ranking}.
Evidently, there is a strong desire to find accurate, topic-specific content in understandable language.
In that regard, it appears almost irrelevant if the media in question previously had significant reach or publicity, so long as they are accurate and provide their sources.
As such, we proceeded to implement an infrastructure meeting these needs.

\subsection{Implementation}
Wikidata~\cite{vrandecic_wikidata_2014} is conceptually capable of meeting most of the requested features, especially if combined with a capable long-text annotation platform such as Wikisource.
However, both of these existing platforms are already operating at capacity while adressing just notable, public domain content.
This was confirmed by several developers from inside and outside the Wikimedia team, including representatives from the Wikibase.cloud team, at the aforementioned MediaWiki Conference.
Hence, we extended the approach from MiMoText~\cite{schoch_smart_2022} by integrating our \nameref{sec:linked_data_wiki} into the Wikibase Ecosystem while expanding to a \nameref{sec:full_tex_wiki}, where legally permitted.
This open foundation was built upon by several microservices, such as a search page, as well as content creation and aggregation automation.
All implemented services are illustrated in detail in the master thesis underlying this article \cite{stehr_digitale_2025} and are available online\footnote{Web-Page: \url{https://github.com/xEatos/dashboardduck}}\footnote{Media Search Service: \url{https://github.com/xEatos/searchsnail}}\footnote{Import \& Integration Service: \url{https://github.com/xEatos/integrationindri}}.

\paragraph{Linked Data Wiki\label{sec:linked_data_wiki}}
The primary persistent layer of the infrastructure is a Wikibase installation.
Building on the overview of scientifically relevant media characteristics by Navarrete et al.~\cite{navarrete_closer_2023}, we identified over 200 different metadata qualities.
Examples include 
textual elements (e.g., transcripts, on-screen text) or
instructor behavior (e.g., tone, lexical diversity).
We complemented properties regarding 
FAIR context (e.g., license, accessibility),
involved actors (e.g., moderators, sponsors) or 
sources provided (e.g., on-screen, in description).
It serves as the central access point as already illustrated in \secref{fig:overview}, responding to queries and linking to other data sources, such as the Full Text Wiki.

\paragraph{Full Text Wiki\label{sec:full_tex_wiki}}
The Full Text Wiki fulfills two roles: first, storing long-form content such as video and podcast transcripts with tens of thousands of characters; second, a space not only to search, but also to annotate, improve, translate, semantify, and further process these textual representations.
Ingesting these texts is legally complex and hence automated only under special circumstances.

\paragraph{Content aggregation automation}
When scaling to millions of heterogeneous artifacts, automation must be modular while ensuring content veracity.
For this purpose, we implemented a service prototype that supports ingesting media metadata.
One such example is a dataset provided by the world lecture project\footnote{\url{https://world-lecture-project.org/}}, from which we created over 5,000 video representations with additional data enriched mainly from the YouTube API.
This service is designed primarily to support advanced users, where most users are expected to interface with the infrastructure primarily as viewers, and most likely via the search page. 

\begin{figure}[b!]
    \centering
    \includegraphics[clip,trim={0.5cm 8.6cm 10cm 11cm},width=0.49\linewidth]{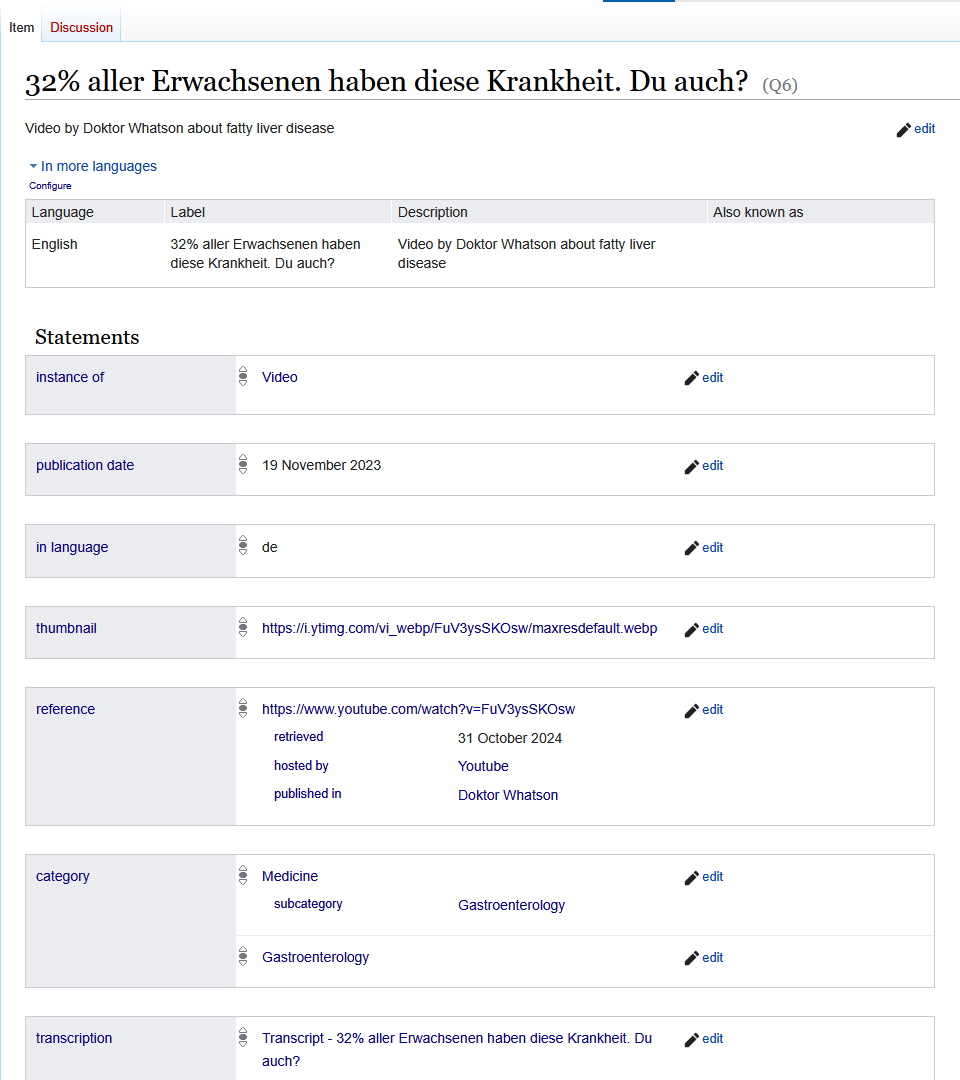}
    \includegraphics[clip,trim={1.5cm 13.5cm 0cm 1.75cm},width=0.49\linewidth]{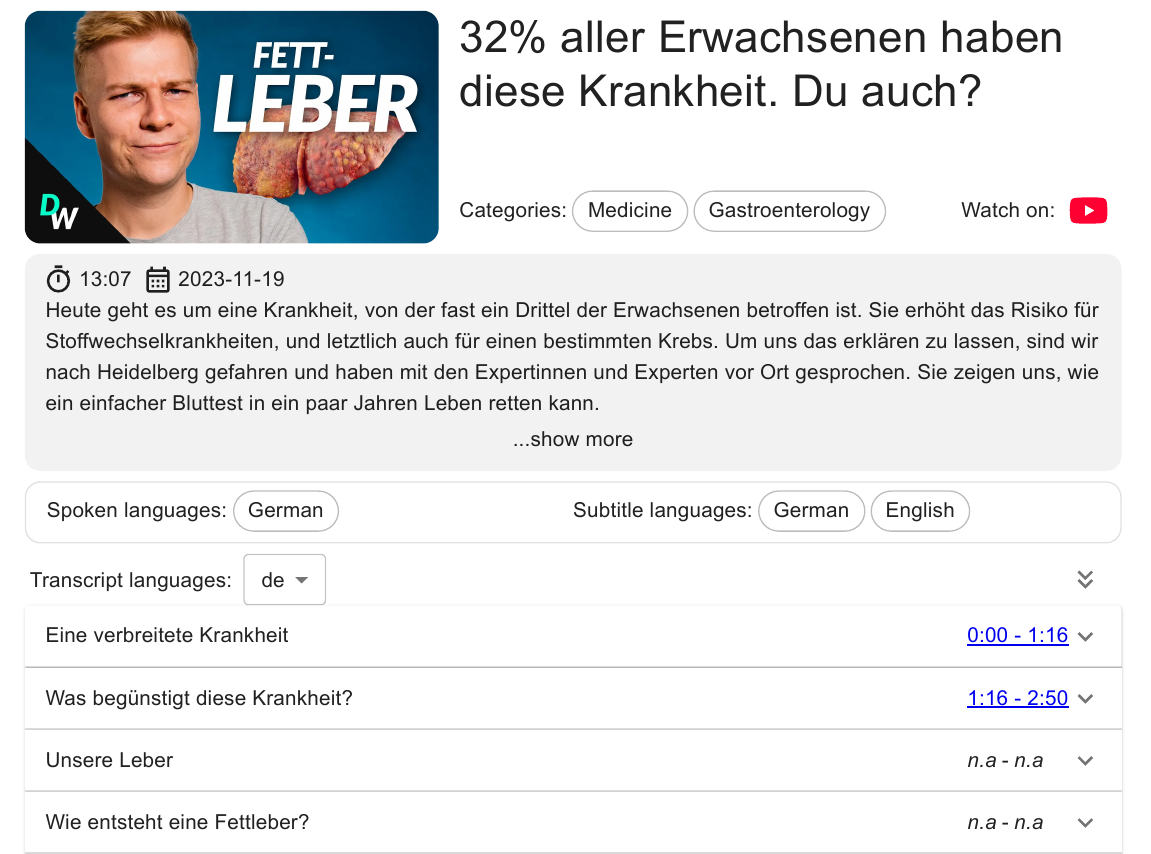}
    \caption{Knowledge graph representation of a media item on wikibase (left), accessed by the Dashboard and displayed as a detail page (right).}
    \label{fig:detail_page}
\end{figure}

\begin{figure}[tb]
    \centering
    \includegraphics[clip,trim={0cm 14.2cm 6.9cm 0cm},width=0.95\linewidth]{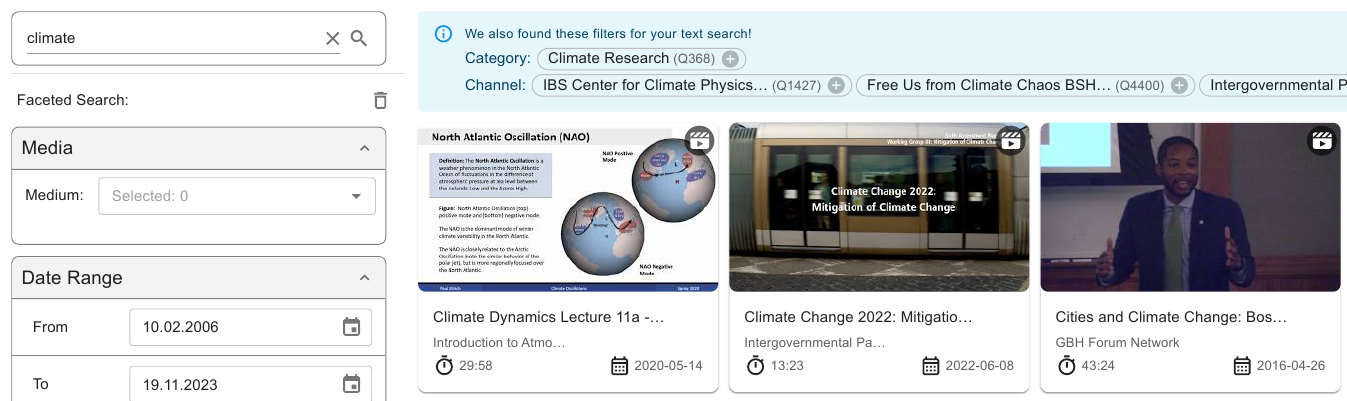}
    \caption{View when searching for climate-related media, incl. filter recommendation.}
    \label{fig:text_search_hint_cutout}
    \vspace{-0.50cm}
\end{figure}

\paragraph{Search page}
The search page is the core interface and entry point for most stakeholders, bringing together all systems.
As a view-only interface, it requires no login, reduces the complexity, and is hence designed to provide accessible usability for any user.
As hinted at in \secref{fig:detail_page} and \ref{fig:text_search_hint_cutout}, properties and content stored in the Wikis can be used to filter for relevant media.

\subsection{Evaluation}
The usability of the search page was examined via an online experiment, evaluating satisfaction, objective and subjective efficiency and effectiveness, and the extent to which the implemented features satisfy the needs.

\paragraph{Design}
Participants were given five tasks designed to test different functionalities and evaluation criteria.
These included searching for: (1) a video by a given title, (2) any video on ``history'' from the University of Göttingen, (3) any video on ``fatty liver'' published after 2022, (4) any video on ``computer science'' which is longer than 60 minutes and published in 2013 or 2014, and (5) any video in English.
Task completion time was recorded to measure objective efficiency, while the number of correctly completed tasks determined effectiveness.
Subjective perceptions of effectiveness, efficiency, and provided support information were captured using an After-Scenario Questionnaire (ASQ) according to Lewis~\cite{lewis_psychometric_1991}, which uses a 7-point Likert scale from 1 (strongly agree) to 7 (strongly disagree).
User satisfaction was measured via the User Experience Questionnaire (UEQ)~\cite{schrepp_construction_2017}, covering attractiveness, perspicuity, efficiency, dependability, stimulation, and novelty.
Participants also rated how well the requested features and criteria  were implemented using a 5-point Likert scale from 1 (very good) to 5 (insufficient).
The study tested four hypotheses:
Users, on average, complete tasks in under three minutes ($H_{A, \text{ efficiency}}$)
, have more than three tasks completed successfully ($H_{A, \text{ effectiveness}}$),
report a $\text{UEQ} > 0.8$ ($H_{A, \text{ UX}}$),
and rate at least four of seven top criteria "good" or better ($H_{A, \text{ criteria}}$).

\begin{table}[b!]
    \vspace{-0.35cm}
    \centering
    \caption{
    Statistical test results. The Shapiro-Wilk test assesses normality $N$ ($W, p$), determining the choice between t-test (for $N=yes$) and Wilcoxon Signed Rank ($N=no$). $t$- \& $p$-values are shown for t-tests; $Z$- and $p$-values for Wilcoxon tests. The null hypothesis ($H_0$) may be rejected based on the Bonferroni-Holm adjusted $p$-value ($p_a$).
    }
    \label{tab:experiment_task_time}
    \resizebox{1\textwidth}{!}{
    \begin{tabular}{|c|ccc|cc|cc|c|c|}
    \hline
    \multirow{2}{*}{Efficiency} & \multicolumn{3}{l|}{Shapiro-Wilk Test} & \multicolumn{2}{c|}{t-Test} & \multicolumn{2}{l|}{\begin{tabular}[c]{@{}l@{}}Wilcoxon \\ Signed Rank\end{tabular}} & \multirow{2}{*}{$p_a$} & \multirow{2}{*}{\begin{tabular}[c]{@{}l@{}}reject \\$H_0$? \end{tabular}} \\ \cline{2-8}
    & \multicolumn{1}{c|}{W} & \multicolumn{1}{c|}{p} & N? & t & p & \multicolumn{1}{c|}{Z} & p & & \\ \hline
    Task 1 & \multicolumn{1}{c|}{0.861} & \multicolumn{1}{c|}{0.0316} & no & \multicolumn{1}{c|}{-} & - & \multicolumn{1}{c|}{-3.265} & 0.001 & 0.005 & yes \\ \hline
    Task 2 & \multicolumn{1}{c|}{0.873} & \multicolumn{1}{c|}{0.046} & no & \multicolumn{1}{c|}{-} & - & \multicolumn{1}{c|}{-2.711} & 0.007 & 0.0336 & yes \\ \hline
    Task 3 & \multicolumn{1}{c|}{0.87} & \multicolumn{1}{c|}{0.0423} & no & \multicolumn{1}{c|}{-} & - & \multicolumn{1}{c|}{-3.668} & $<$ 0.001 & 0.001 & yes \\ \hline
    Task 4 & \multicolumn{1}{c|}{0.68} & \multicolumn{1}{c|}{$<$ 0.001} & no & \multicolumn{1}{c|}{-} & - & \multicolumn{1}{c|}{-1.57} & 0.117 & 0.583 & no \\ \hline
    Task 5 & \multicolumn{1}{c|}{0.905} & \multicolumn{1}{c|}{0.1315} & yes & \multicolumn{1}{c|}{-26.179} & $<$ 0.001 & \multicolumn{1}{c|}{-} & - & $<$ 0.001 & yes \\ \hline
    \end{tabular}
    }
    \label{tbl:effizienz}
    \vspace{-0.65cm}
\end{table}

\begin{figure}[bt]
    \centering
    \includegraphics[width=1.0\linewidth]{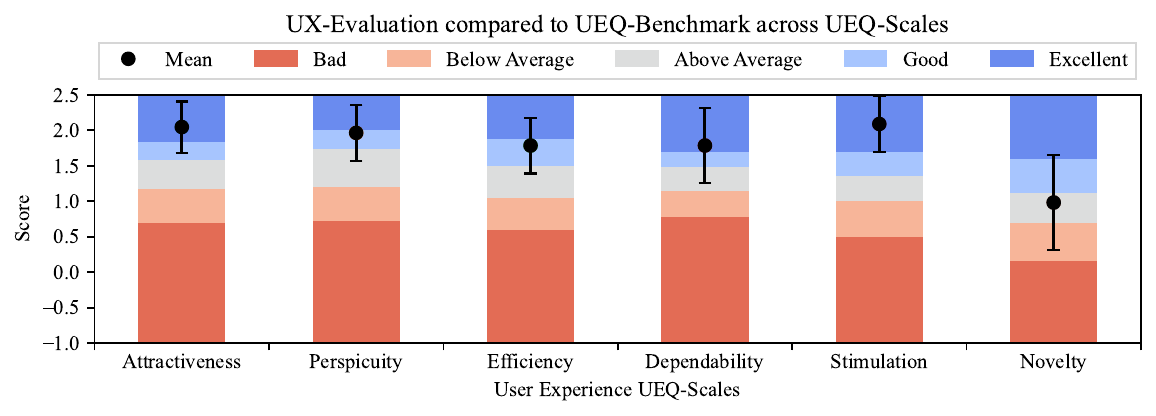}
    \vspace{-0.75cm}
    \caption{UEQ benchmark results across six UX scales (Attractiveness, Perspicuity, Efficiency, Dependability, Stimulation, and Novelty). The colored bands represent qualitative rating categories ("Bad" to "Excellent"). The black dots indicate the mean scores for each scale, the whiskers indicate the $95\%$ confidence intervals for the scores.}
    \label{fig:benchmark}
    \vspace{-0.25cm}
\end{figure}

\paragraph{Results}
\secref{tab:experiment_task_time} details the statistical tests for the hypotheses, confirming showing positive objective efficiency ($H_{A, \text{ efficiency}}$) and effectiveness ($H_{A, \text{ effectiveness}}$).
\secref{fig:benchmark} illustrates the equally positive usability results ($\overline{{UEQ}} = 1.78$ ($H_{A, \text{ UX}}$), namely three excellent, two good, and one above-average ratings, compared to the benchmark~\cite{schrepp_construction_2017}. 
The ASQ results were equally positive, with a mean effectiveness of 2.04 ($\sigma{} = 0.76$), efficiency of 1.67 ($\sigma{} = 0.68$), and provided support information of 2.075 ($\sigma{} = 0.64$).
Participants rated three of the seven highest-priority criteria as “good” or better, narrowly missing the $H_{A, \text{ criteria}}$ threshold of four.
While the current SciCom Wiki system does not yet fully satisfy all aspired features, it provides a consistently well-rated usable and conceptually extensible foundation for future work.
Beyond supporting search, our FAIR data library also systematically establishes the basis for large-scale processing of not-(yet-)textual media, facilitating another highly requested feature: fact-checking.

\section{Computational Fact-Checking of Online Discourse\label{sec:fact-checking}}
We have evaluated the computational fact-checking capacities by developing and evaluating a neurosymbolic-AI supported framework\footnote{\url{https://github.com/cTremel/Scientific-Knowledge-fit-for-Society}} for a climate communication use case. 
The approach and tool were refined using insights gathered in two interim presentations, complemented with ten one-on-one expert interviews and a user survey, as detailed in the master thesis by Tremel~\cite{tremel_scientific_2024}.

\subsection{Approach}
\begin{figure}[bt]
    \vspace{-0.5cm}
    \centering
    \includegraphics[width=1.0\linewidth]{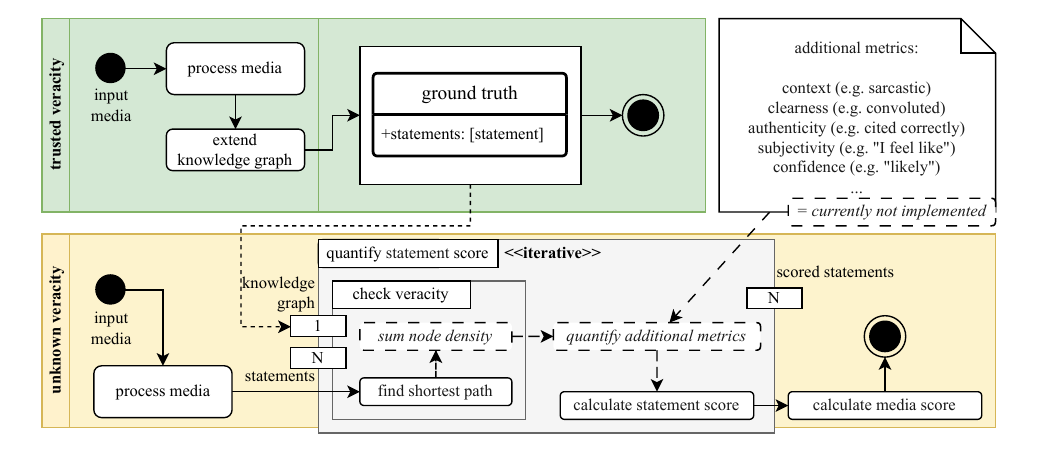}
    \caption{Proposed scoring pipeline consisting of (i) Trusted statements extend the ground truth knowledge graph, (ii) untrusted are checked for veracity using graph analysis on the ground truth, concluding in (iii) a final score calculation.}
    \label{fig:pipeline}
\end{figure}
\begin{figure}[bt]
    \centering
    \includegraphics[clip,trim={0.70cm 0.50cm 0.70cm 0.2cm},width=1.0\linewidth]{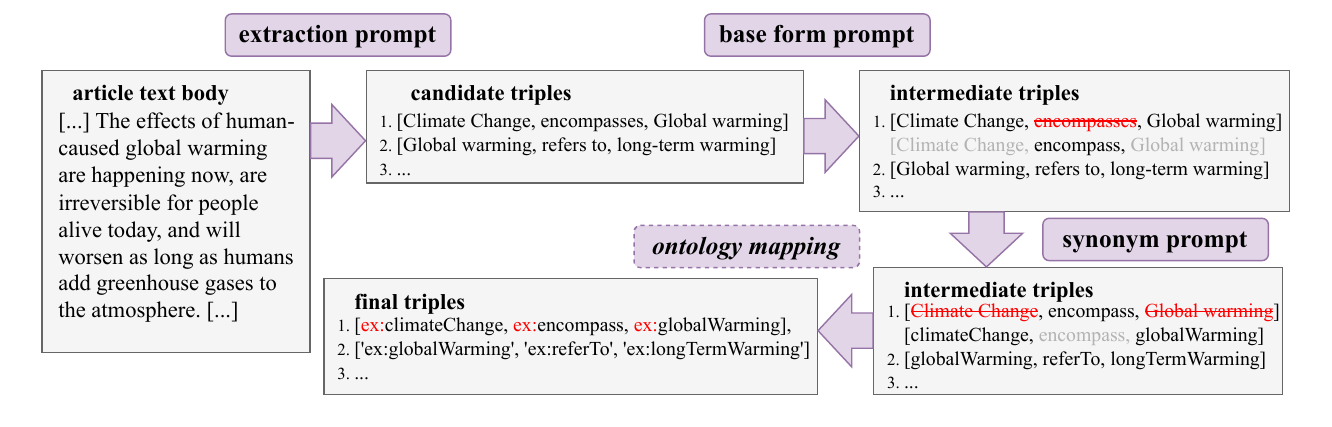}
    \caption[Triple extraction workflow example from text body to aligned triples. An LLM was used to handle initial extraction, base forms, and synonyms. A prototypical ontology mapping was implemented to a placeholder example ontology.]{Triple extraction workflow example from text body to aligned triples. An LLM was used to handle initial extraction, base forms, and synonyms. A prototypical ontology mapping was implemented to a placeholder example ontology.}
    \label{fig:CFCoOD_example}
\end{figure}
Our goal is to improve the quality of public information exchange using a scientific accuracy score by computationally evaluating media statements based on expert research.
As motivated in \secref{sec:related}, this requires a solid knowledge base built on statements from peer-reviewed scientific literature, structured in a knowledge graph.
\secref{fig:pipeline} shows that first, trusted source media needs to be processed, transformed into semantic statements (unless already available), aligned with a cohesive data model, and ingested into a knowledge graph.
In our case, the ground truth utilized consisted of headline statements of the IPCC AR6~\cite{lee_ipcc_2023}, the highly synthesized consensus on climate science.
Only then can media of unknown veracity, such as viral posts or videos, be processed and their statements be subsequently aligned and compared with the ground truth knowledge graph.
In \secref{fig:CFCoOD_example}, this workflow is illustrated using an example from an online article text body and finishing with aligned triples.
The following subsections and paragraphs describe this process in detail.

\subsubsection{Process media}
\begin{figure}[tb]
    \centering
    \includegraphics[clip,trim={0.5cm 0.25cm 0.75cm 0.25cm},width=1.0\linewidth]{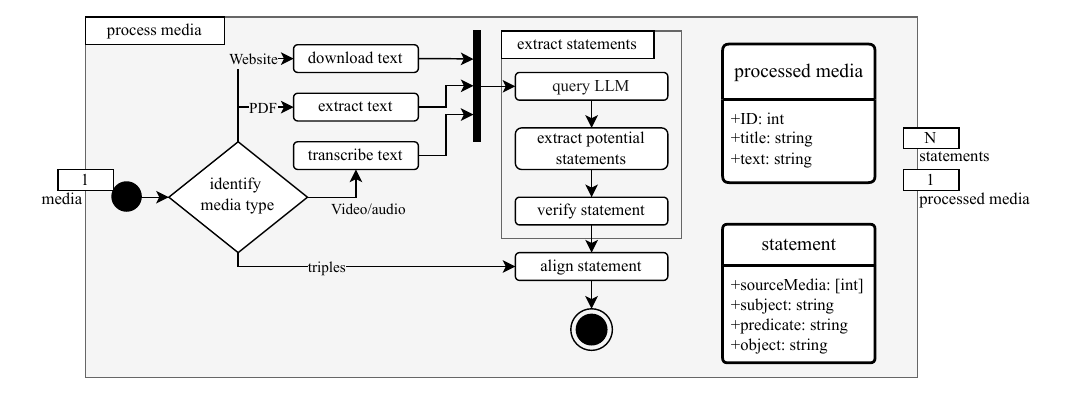}
    \caption{Description of processing media consisting of (i) textualizing different media file types, (ii) LLM-based statement extraction, verification, and alignment.}
    \label{fig:process}
\end{figure}

The type of media defines the required processing as depicted in \secref{fig:process}. 
Text from web documents is extracted using the Python library Beautiful Soup, PDF files using PDFMiner, and Audio or video files are transcribed using OpenAI Whisper.
Text extracted this way may lose structural context information or be distorted with noise, such as headings and editorial information.
If the media is not already described via triple statements (e.g., RDF dump), triples are extracted in the next step.

\paragraph{Statement extraction}
We investigated existing techniques to efficiently extract reliable triple statements at scale, considering context, particularly for the climate domain, but preferably domain-independent. 
Since our research could not identify an approach that suited our needs, we tested three different approaches.
First, semantic role labeling (SRL)~\cite{he_deep_2017} was used to create Abstract Meaning Representation (AMR)~\cite{banarescu_abstract_2013} graphs.
Disambiguating unrestricted AMR graphs and using them for semantically consistent and reliable fact-checking was highly time-intensive.
Once set up, these graphs work consistently, but their setup does not scale well to the complexity and number of statements required.
Second, we explored triple extraction using a domain-specific Named Entity Recognition (NER) model.
Training a domain-specific model combined with an ontology could provide better results, but is cost-intensive, time-consuming, and not very suitable for the heterogeneous domains of public discourse.
Our third approach utilized Large Language Models (LLMs) to extract and align triples.
While LLMs are currently not reliable either, human-in-the-loop supervision promises faster, sufficiently reliable results.
After iterative testing, these initial assessments were reviewed by NLP and structured knowledge experts.
The findings were substantiated, and the LLM-based approach was deemed the best fit for the scope, potentially to be complemented with a NER-aided ontology-based mediator.

\paragraph{Alignment} Predicate normalization and semantic alignment, such as synonym detection, concluded the triple processing.
Each triple is assigned a media ID array, allowing the statement origin to be traceable through multiple sources.

\paragraph{Extension\label{sec:extendKG}}
If the source was trusted (i.e., IPCC), we ingested the RDF serialized triples into a GraphDB database.
Otherwise, the media progressed to \nameref{sec:check}.
While not included in the final version, we experimented with using an ontology to help with inferring information and detecting inconsistent triples.
Potentially, federating multiple graphs 
and sources with different publication dates
would allow for a broader, up-to-date observation of scientific knowledge.
 
\subsubsection{Veracity Checking\label{sec:check}}
\begin{figure}[bt]
    \centering    \includegraphics[width=\linewidth]{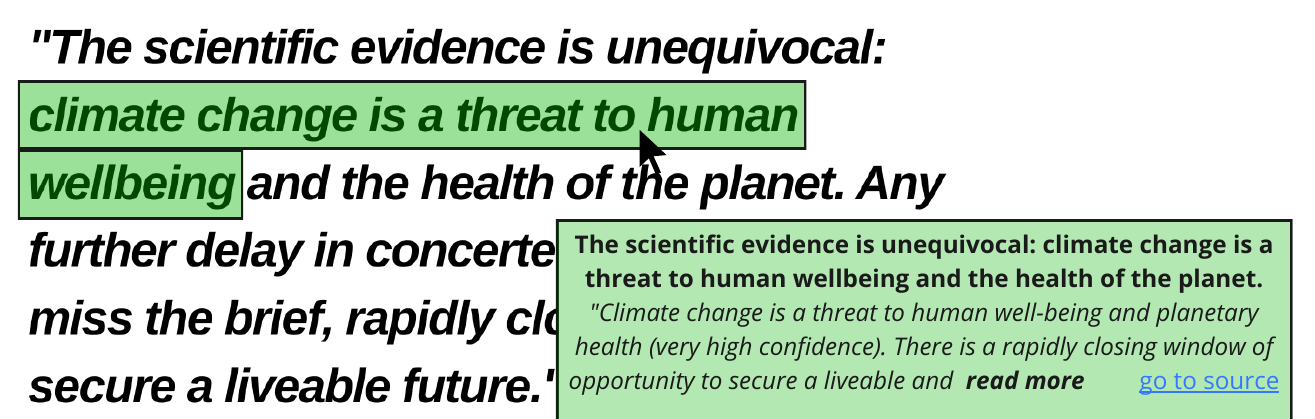}
    \caption[User interface mock-up showing the statement veracity score with color coding and ground truth reference.]{A User interface mock-up representing the statement\footnotemark{} veracity score via positive (green) color coding and providing a ground truth reference as explanation.}
    \label{fig:UI}
\end{figure}

The veracity check involves knowledge graph analysis against the ground truth, initially searching for an exact match.
If no match is found, a path check according to 
Ciampaglia et al.~\cite{ciampaglia_computational_2015} can approximate the veracity,
which should only be interpreted as an indication, not a sufficient check.
\secref{fig:UI} displays a user interface mock-up, indicating the veracity of a statement by using color coding and referring to source information from the ground truth.

\paragraph{Quantify additional metrics}
The annotation quality of the accuracy score can be improved by considering not only veracity, but also temporal relevance, confidence, clearness, transparency, information depth, objectivity, and rationality.
\footnotetext{Example NASA Article \url{https://science.nasa.gov/climate-change/effects/}}
The overall accuracy score $s_{acc} \in [0, 1]$ is calculated using the weighted individual metrics: $s_{acc} = \sum_{i=1}^{n} s_i \cdot w_i$, where $s_i \in [0, 1]$ and $\sum_{i=1}^{n} w_i = 1$. 
According to our investigation of computational fact-checking, accuracy consists mostly of veracity: $w_{ver} \geqslant 0.5 \geqslant \sum_{i=1, \ i \neq ver}^{n} w_i$. 
Since no satisfactory computational quantifications could be identified for the additional metrics, our accuracy scoring is currently limited to veracity ($w_{ver} = 1$).

\subsection{Evaluation}
To ensure that the methods and tools are up-to-date and properly contextualized, the approach and its implementation were evaluated through expert feedback during several presentations and interviews with knowledge representation and processing experts from the ORKG team.
Additionally, a user survey was conducted to assess the current state and inform future development.

\paragraph{Expert Interviews} Ten experts provided detailed responses in individual semi-structured interviews. 
In sessions averaging 28 min (total: 4:42:00, max: 0:43:55, min: 0:13:39, $\sigma$: 0:10:40), the following insights were gathered:
\begin{itemize}
    \item The system is suitable as a personal assistant and should focus on the interoperability of tools ranging from trusted graphs to browsers to PDF readers.
    \item Knowledge graphs are state-of-the-art interoperable knowledge representations. When no exact match is found, embeddings or graph walks should be utilized, particularly with increased weighting of incoming edges.
    \item While LLMs are state-of-the-art for scalable statement extraction, they are unreliable for semantic parsing and require the identification of non-reproducible or hallucinated triples.
    \item A NER mediator could complement the LLM to ensure consistency in moderating statements, instances, and classes.
\end{itemize}

\begin{figure}[bth!]
    \vspace{-0.5cm}
    
    \includegraphics[width=.95\textwidth]{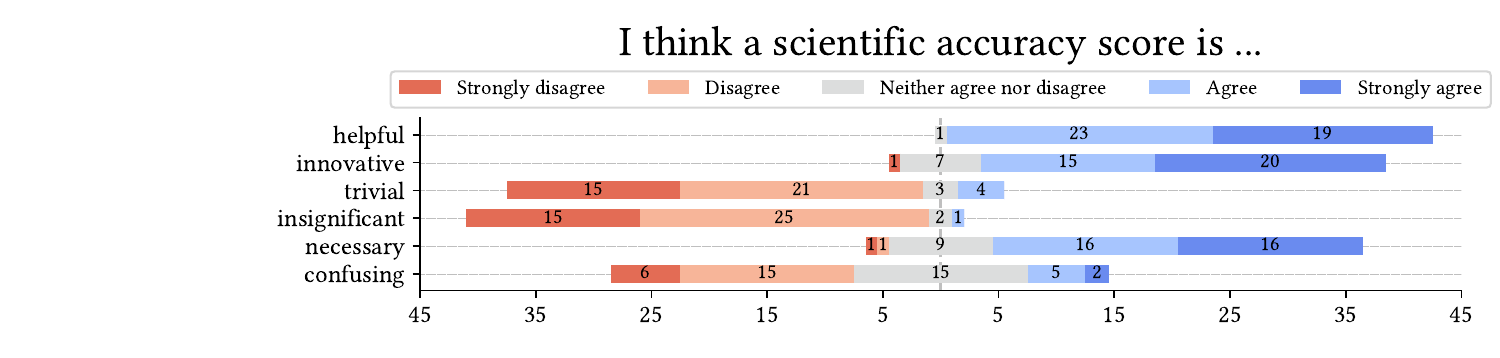}

    \vspace{-0.2cm}
    
    \includegraphics[width=.95\textwidth]{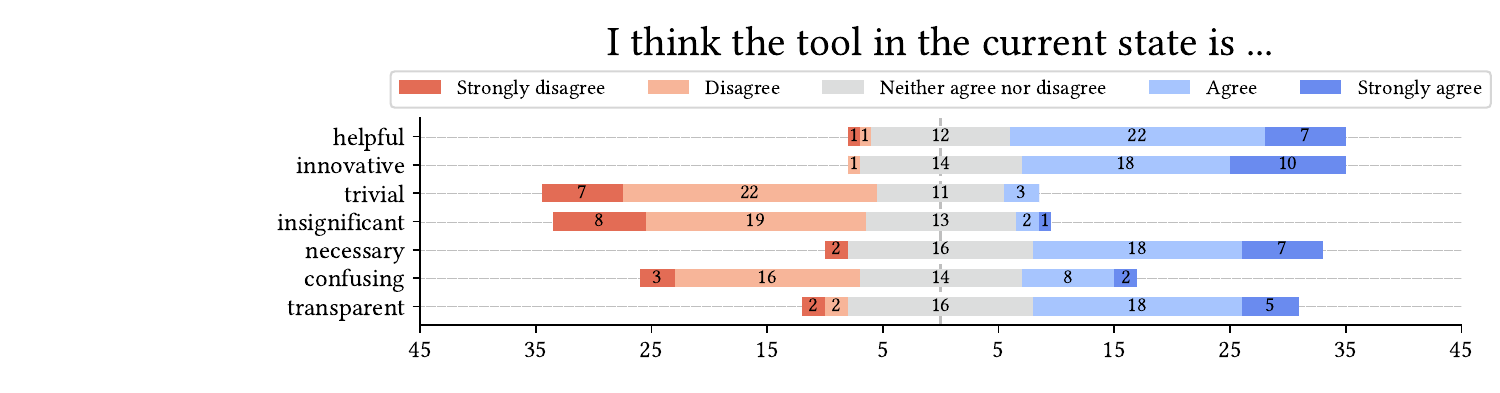}
    
    \vspace{-0.2cm}
    
    \includegraphics[width=.95\textwidth]{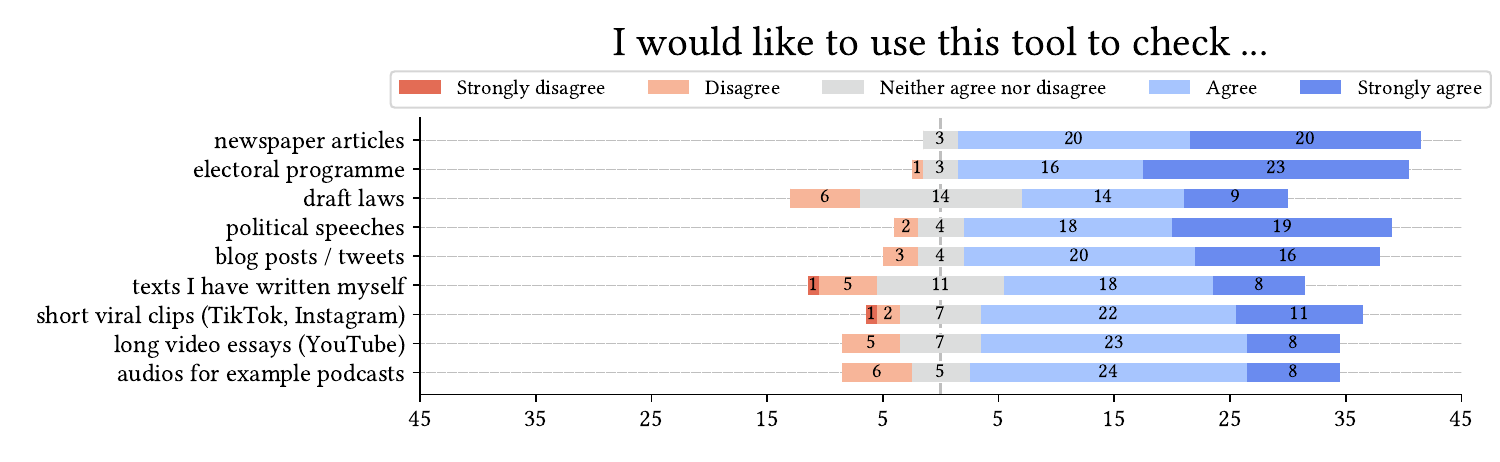}
    
    \vspace{-0.2cm}
    
    \includegraphics[width=.95\textwidth]{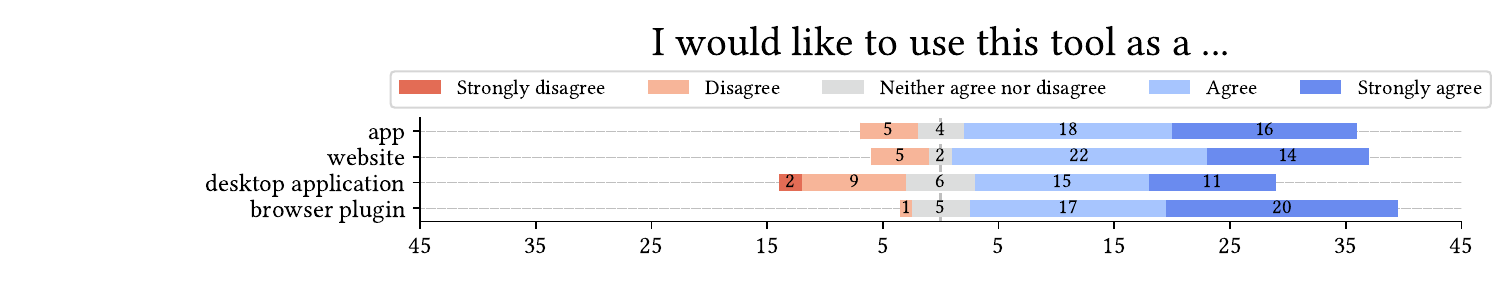}
    
    \vspace{-0.2cm}
    
    \caption{Results of 43 participants 
    assessing the demand for a scientific accuracy score, approving the tool in its current state, and indicating suitability for various media.}
    
    \vspace{-0.75cm}
    
    \label{fig:user}
\end{figure}

\paragraph{User Survey}
The survey was distributed digitally with  broad inclusion criteria, as all consumers of online content are potential users.
It presented the tool and inquired the users evaluation of it with several 5-point Likert scale questions, as depicted in \secref{fig:user}.
The closing demographic questions revealed that the 43 anonymous participants were primarily in the age group 19-31 and tended to have some educational degree (high school, bachelor or master).
The results indicate the necessity and usability of the tool.
Notably, participants expressed interest in fact-checking a wide range of media formats, further supporting the demand for a versatile and scalable infrastructure in science communication.

\section{Discussion and Future Work\label{sec:discussion}}
We demonstrated systems to support a SciCom KI in the most requested identified features, such as fact-checking, structuring, and filtering media.
FAIR representation of video and podcast information enables collaboration, post-processing, and veracity assessment, improving navigation for researchers and viewers alike.
While promising in structurally highlighting reliable knowledge and systematically addressing misinformation, our work revealed several issues.

Science communication media is designed for human readability, not machine readability, yet machines facilitate its distribution and digestion.
Extensive manual fact-checking resources exist, but are largely not available as high-quality FAIR data, severely limiting the computational fact-checking potential.
Especially for urgent domains such as climate change, larger-scale semantification collaboration is required to build a comprehensive corpus to check against.
The SciCom KI requires collaborative, FAIR systems to meet its needs in an increasingly non-textual, factually ambiguous media landscape.

Yet, the analysis our in-person experiments suggests that general users seem to prefer systems to abstract away the underlying complexity of knowledge graphs and simplify the interaction with them.
While structured semantic data aids precision, it does not inherently improve accessibility for a broad audience.
Therefore, presenting this information in a user-friendly and easily digestible way remains a critical challenge.
A digital library with an integrated fact-checking system alone is not enough to stand out in an environment full of unreliable information.
To meaningfully contribute to public discourse, the SciCom Wiki needs to integrate current media and enable their analysis in an FAIR, yet efficient and scalable manner.

Our proposed, demonstrated, and evaluated systems can meet several of these requirements, yet they too require systematic and individual support.

SciCom KI, like any KI, consists of people, artifacts, and institutions, all of which require different approaches to unify into a cohesive network.
Generating, sharing, and maintaining FAIR knowledge equally require individual approaches and support from each actor mentioned above.
As such, our system artifacts to process, maintain and share FAIR data require people and institutions that realize their potential.
Technologically, the SciCom KI potential vastly exceeds its actual state-of-the-art.
Both systems have shown that plenty of knowledge is generated and readily human-readable, yet to meet the SciCom needs, this knowledge, alongside its creators, needs to be integrated into a robust network.

If future work implements systems like we proposed and enacts them in a SciCom KI incorporating people and institutions, this FAIR knowledge infrastructure could fill an important gap towards a resilient, fact-based democratic society.
Just as the introduction of Wikipedia and Wikidata has reformed how we seek, process, and exchange information, alongside unintended consequences such as providing presumably the largest structured knowledge base for modern LLMs, this next step could reform non-textual knowledge exchange.
Our findings indicate that even fully implementing all proposed changes will not solve the misinformation crisis.
An exclusive focus on factuality risks missing other critical dimensions, such as context (e.g., sarcastic), clearness (e.g., convoluted), authenticity (e.g., cited incorrectly), subjectivity (e.g., ``I feel like''), or confidence (e.g., ``likely'').
Even statements that are not factually false, including subjective opinions, can strongly shape discourse.
Similarly, a factually correct statement taken out of context can misrepresent a complex topic.
Citizens who distrust scientists will also not accept a ground truth built upon science.
Reaching the intended audience always remains a challenging issue in the attention economy.
All these issues need to be individually addressed, beyond any single KI.
Our proposed SciCom Wiki particularly fills the gap of a repository for scalable crowd-sourcing of FAIR ground truth synthesis and extensive popular media processing.
We recommend advancing the SciCom KI, including already active actors, furthering effective result dissemination, and dedicated resources for science skeptics.

\section{Conclusion\label{sec:conclusion}}
This work illustrated the problem of FAIR media processing in facilitating a Science Communication Knowledge Infrastructure (SciCom KI).
Particularly usable and scalable approaches for non-textual media storage, utilization, and computational fact-checking have been designed, demonstrated, and evaluated as usable solutions to assist the effort against the (mis-)information flood.
While plenty of fact-checking and ground-truth resources exist, too few are as machine-readable as the literature recommends.
Equally, apparently no existing FAIR digital library scales to the rapid creation and dissemination of civically central information media, particularly videos and podcasts.
Following Wikidata's Open Linked Data Approach, our approach has been shown to effectively address both of these issues while synergistically benefiting from each other and an overarching, proposed FAIR SciCom KI.
The individual solutions have been evaluated in respective surveys and interviews. 
The evaluation of our digital library system surveyed 53 participants and interviewed another 11 for SciCom KI requirements and evaluated them against these with 14 stakeholders.
The neurosymbolic fact-checking approach followed an iterative development approach, integrating the feedback of 10 experts into its development, and evaluated the final results against another 43 survey participants and 10 individual expert interviews.
Our systems contribute to solving their respective challenges, yet further scaling, adoption, and expansion are necessary to develop the SciCom KI and check misinformation adequately.
The SciCom Wiki demonstrated that FAIR, collaborative, and user-friendly infrastructure for non-textual media is feasible, well-received by users, and lays the groundwork for more transparent, verifiable, and accessible science communication.

\begin{credits}
\subsubsection{\ackname} 
This work was co-funded by the DFG SE2A Excellence Cluster, as well as the NFDI4Ing project funded by the German Research Foundation (project number 442146713) and NFDI4DataScience (project number 460234259).

\subsubsection{\discintname}
The authors have no competing interests to declare that are relevant to the content of this article.
\end{credits}
%
%
%
\bibliographystyle{splncs04}
\bibliography{bibliography}

\begin{thebibliography}{10}
\providecommand{\url}[1]{\texttt{#1}}
\providecommand{\urlprefix}{URL }
\providecommand{\doi}[1]{https://doi.org/#1}

\bibitem{domo}
Domo {Resource} - {Data} {Never} {Sleeps} 11.0 (2023), \url{https://www.domo.com/learn/infographic/data-never-sleeps-11}

\bibitem{auer_improving_2020}
Auer, S., Oelen, A., Haris, M., Stocker, M., D’Souza, J., et~al.: Improving {Access} to {Scientific} {Literature} with {Knowledge} {Graphs}. Bibliothek Forschung und Praxis  \textbf{44}(3),  516--529 (Dec 2020). \doi{10.1515/bfp-2020-2042}, publisher: De Gruyter

\bibitem{banarescu_abstract_2013}
Banarescu, L., Bonial, C., Cai, S., other: Abstract meaning representation for sembanking. In: Proceedings of the 7th Linguistic Annotation Workshop and Interoperability with Discourse. pp. 178--186. Association for Computational Linguistics (2013), \url{https://aclanthology.org/W13-2322}

\bibitem{brennen_types_2020}
Brennen, J.S., Simon, F.M., Howard, P.N., Nielsen, R.K.: Types, sources, and claims of {COVID}-19 misinformation  (2020), \url{https://ora.ox.ac.uk/objects/uuid:178db677-fa8b-491d-beda-4bacdc9d7069}, publisher: Reuters Institute for the Study of Journalism

\bibitem{burns_science_2003}
Burns, T.W., O'Connor, D.J., Stocklmayer, S.M.: Science {Communication}: {A} {Contemporary} {Definition}. Public Understanding of Science  \textbf{12}(2),  183--202 (Apr 2003). \doi{10.1177/09636625030122004}, publisher: SAGE Publications Ltd

\bibitem{lee_ipcc_2023}
Calvin, K., Dasgupta, D., Krinner, G., et~al.: {IPCC}, 2023: Full report [core writing team, h. lee and j. romero (eds.)]. {IPCC}, geneva, switzerland. (2023). \doi{10.59327/IPCC/AR6-9789291691647}, \url{https://www.ipcc.ch/report/ar6/syr/}, edition: First

\bibitem{ciampaglia_computational_2015}
Ciampaglia, G.L., Shiralkar, P., Rocha, L.M., other: Computational fact checking from knowledge networks  \textbf{10}(6),  e0128193 (2015). \doi{10.1371/journal.pone.0128193}, publisher: Public Library of Science

\bibitem{cook2013}
Cook, J., Nuccitelli, D., Green, S.A., Richardson, M., Winkler, B., et~al.: Quantifying the consensus on anthropogenic global warming in the scientific literature. Environmental Research Letters  \textbf{8}(2),  024024 (Jun 2013). \doi{10.1088/1748-9326/8/2/024024}

\bibitem{davarpanah_climate_2023}
Davarpanah, A., Babaie, H.A., Huang, G.: Climate system ontology: A formal specification of the complex climate system. In: Latest Advances and New Visions of Ontology in Information Science. {IntechOpen} (2023). \doi{10.5772/intechopen.110809}

\bibitem{dessi_scicero_2022}
Dessí, D., Osborne, F., Reforgiato~Recupero, D., Buscaldi, D., Motta, E.: {SCICERO}: A deep learning and {NLP} approach for generating scientific knowledge graphs in the computer science domain  \textbf{258},  109945 (2022). \doi{10.1016/j.knosys.2022.109945}

\bibitem{10.5555/1805940}
Edwards, P.N.: A Vast Machine: Computer Models, Climate Data, and the Politics of Global Warming. The MIT Press (2010)

\bibitem{emissiongapReport2023}
Environment, U.N.: Emissions {Gap} {Report} 2023 (Aug 2023), \url{http://www.unep.org/resources/emissions-gap-report-2023}, section: publications

\bibitem{fahnrich_exploring_2023}
Fähnrich, B., Weitkamp, E., Kupper, J.F.: Exploring ‘quality’ in science communication online: {Expert} thoughts on how to assess and promote science communication quality in digital media contexts. Public Understanding of Science  \textbf{32}(5),  605--621 (Jul 2023). \doi{10.1177/09636625221148054}, publisher: SAGE Publications Ltd

\bibitem{giles_internet_2005}
Giles, J.: Internet encyclopaedias go head to head. Nature  \textbf{438}(7070),  900--901 (Dec 2005). \doi{10.1038/438900a}, publisher: Nature Publishing Group

\bibitem{hagenhoff_neue_2007}
Hagenhoff, S., Seidenfaden, L., Ortelbach, B., Schumann, M.: Neue {Formen} der {Wissenschaftskommunikation}: eine {Fallstudienuntersuchung}. Universitätsverlag Göttingen (2007). \doi{10.17875/gup2007-208}, accepted: 2020-04-15T02:34:09Z

\bibitem{he_deep_2017}
He, L., Lee, K., Lewis, M., Zettlemoyer, L.: Deep semantic role labeling: What works and what's next. In: Proceedings of the 55th Annual Meeting of the Association for Computational Linguistics (Volume 1: Long Papers). pp. 473--483. Association for Computational Linguistics (2017). \doi{10.18653/v1/P17-1044}

\bibitem{islam_knowurenvironment_2022}
Islam, S., Proma, A., Zhou, Y., Akter, S.N., Wohn, C., Hoque, E.: {KnowUREnvironment}: An automated knowledge graph for climate change and environmental issues  (Nov 2022), \url{https://www.climatechange.ai/papers/aaaifss2022/3}

\bibitem{kappel_why_2019}
Kappel, K., Holmen, S.J.: Why {Science} {Communication}, and {Does} {It} {Work}? {A} {Taxonomy} of {Science} {Communication} {Aims} and a {Survey} of the {Empirical} {Evidence}. Frontiers in Communication  \textbf{4} (Oct 2019). \doi{10.3389/fcomm.2019.00055}, publisher: Frontiers

\bibitem{karasti_infrastructure_2010}
Karasti, H., Baker, K.S., Millerand, F.: Infrastructure {Time}: {Long}-term {Matters} in {Collaborative} {Development}. Computer Supported Cooperative Work (CSCW)  \textbf{19}(3),  377--415 (Aug 2010). \doi{10.1007/s10606-010-9113-z}

\bibitem{karasti_knowledge_2016}
Karasti, H., Millerand, F., Hine, C.M., Bowker, G.C.: Knowledge infrastructures: {Part} {I}. Science \& Technology Studies  \textbf{29}(1),  2--12 (Feb 2016). \doi{10.23987/sts.55406}, number: 1

\bibitem{kulczycki_transformation_2013}
Kulczycki, E.: Transformation of {Science} {Communication} in the {Age} of {Social} {Media}. Teorie vědy / Theory of Science  \textbf{35}(1),  3--28 (May 2013). \doi{10.46938/tv.2013.172}

\bibitem{lewis_psychometric_1991}
Lewis, J.R.: Psychometric evaluation of an after-scenario questionnaire for computer usability studies: the {ASQ}. SIGCHI Bull.  \textbf{23}(1),  78--81 (Jan 1991). \doi{10.1145/122672.122692}

\bibitem{mackenzie_science_2019}
MacKenzie, L.E.: Science podcasts: analysis of global production and output from 2004 to 2018. Royal Society Open Science  (Jan 2019). \doi{10.1098/rsos.180932}, publisher: The Royal Society Publishing

\bibitem{marin_arraiza_tibav_2015}
Marín~Arraiza, P., Strobel, S.: The {TIB}{\textbar}{AV} {Portal} as a {Future} {Linked} {Media} {Ecosystem}. In: Proceedings of the 24th {International} {Conference} on {World} {Wide} {Web}. pp. 733--734. {WWW} '15 {Companion}, Association for Computing Machinery, New York, NY, USA (May 2015). \doi{10.1145/2740908.2742912}

\bibitem{navarrete_closer_2023}
Navarrete, E., Nehring, A., Schanze, S., Ewerth, R., Hoppe, A.: A {Closer} {Look} into {Recent} {Video}-based {Learning} {Research}: {A} {Comprehensive} {Review} of {Video} {Characteristics}, {Tools}, {Technologies}, and {Learning} {Effectiveness} (Aug 2023). \doi{10.48550/arXiv.2301.13617}, arXiv:2301.13617 [cs]

\bibitem{schrepp_construction_2017}
Schrepp, M., Hinderks, A., Thomaschewski, J.: Construction of a {Benchmark} for the {User} {Experience} {Questionnaire} ({UEQ}). International Journal of Interactive Multimedia and Artificial Intelligence  \textbf{4}(4),  40--44 (2017). \doi{10.25968/opus-3397}, publisher: Universidad Internacional de La Rioja

\bibitem{schoch_smart_2022}
Schöch, C., Hinzmann, M., Röttgermann, J., Dietz, K., Klee, A.: Smart {Modelling} for {Literary} {History}. International Journal of Humanities and Arts Computing  \textbf{16}(1),  78--93 (Mar 2022). \doi{10.3366/ijhac.2022.0278}, publisher: Edinburgh University Press

\bibitem{stehr_digitale_2025}
Stehr, N.: Eine digitale {Wissensinfrastruktur} zur {Bereitstellung} von {Informationen} über wissenschaftliche {Videos} und {Podcasts}  (Apr 2025). \doi{10.15488/18996}, publisher: Hannover : Gottfried Wilhelm Leibniz Universität

\bibitem{statista_daten_2023}
Tenzer, F.: Daten - {Volumen} der weltweit generierten {Daten} bis 2027 (May 2023), \url{https://de.statista.com/statistik/daten/studie/267974/umfrage/prognose-zum-weltweit-generierten-datenvolumen/}, data from IDC Global DataSphere

\bibitem{thorne_fever_2018}
Thorne, J., Vlachos, A., Christodoulopoulos, C., Mittal, A.: {FEVER}: a {Large}-scale {Dataset} for {Fact} {Extraction} and {VERification}. In: Walker, M., Ji, H., Stent, A. (eds.) Proceedings of the 2018 {Conference} of the {North} {American} {Chapter} of the {Association} for {Computational} {Linguistics}: {Human} {Language} {Technologies}, {Volume} 1 ({Long} {Papers}). pp. 809--819. Association for Computational Linguistics, New Orleans, Louisiana (Jun 2018). \doi{10.18653/v1/N18-1074}

\bibitem{tremel_scientific_2024}
Tremel, C.S.: Scientific {Knowledge} fit for society - {Scoring} scientific accuracy in climate change related news articles. Master's thesis, Hannover : Gottfried Wilhelm Leibniz Universität (2024). \doi{10.15488/17173}

\bibitem{varvantakis_wikibase_2025}
Varvantakis, C.: The {Wikibase} {Software}: {Data} and {Collections} in the {Linked} {Open} {Data} {Web} (Jan 2025). \doi{10.5281/zenodo.14655779}

\bibitem{vrandecic_wikidata_2014}
Vrandečić, D., Krötzsch, M.: Wikidata: a free collaborative knowledgebase. Commun. ACM  \textbf{57}(10),  78--85 (Sep 2014). \doi{10.1145/2629489}

\bibitem{wilkinson_fair_2016}
Wilkinson, M.D., Dumontier, M., Aalbersberg, I.J., Appleton, G., Axton, M., Baak, A., Blomberg, N., Boiten, J.W., da~Silva~Santos, L.B., Bourne, P.E., Bouwman, J., Brookes, A.J., Clark, T., Crosas, M., Dillo, I., Dumon, O., Edmunds, S., Evelo, C.T., Finkers, R., Gonzalez-Beltran, A., Gray, A.J.G., Groth, P., Goble, C., Grethe, J.S., Heringa, J., ’t Hoen, P.A.C., Hooft, R., Kuhn, T., Kok, R., Kok, J., Lusher, S.J., Martone, M.E., Mons, A., Packer, A.L., Persson, B., Rocca-Serra, P., Roos, M., van Schaik, R., Sansone, S.A., Schultes, E., Sengstag, T., Slater, T., Strawn, G., Swertz, M.A., Thompson, M., van~der Lei, J., van Mulligen, E., Velterop, J., Waagmeester, A., Wittenburg, P., Wolstencroft, K., Zhao, J., Mons, B.: The {FAIR} {Guiding} {Principles} for scientific data management and stewardship. Scientific Data  \textbf{3}(1),  160018 (Mar 2016). \doi{10.1038/sdata.2016.18}, publisher: Nature Publishing Group

\bibitem{wittenborg_scicom_2024}
Wittenborg, T.: {SciCom} {Wiki} [{EN}] - {FAIR} knowledge infrastructure for educational content {\textbar} {SMWCon} {Fall} 2024 (2024), \url{https://doi.org/10.5446/69939}

\end{thebibliography}
\end{document}